\documentclass[conference]{IEEEtran}
\IEEEoverridecommandlockouts
\usepackage{booktabs}
\usepackage{cite}
\usepackage{amsmath,amssymb,amsfonts}
\usepackage{algorithmic}
\usepackage{graphicx}
\usepackage{textcomp}
\usepackage{xcolor}
\usepackage{graphicx}
\usepackage{amsmath}
\usepackage{amssymb}
\usepackage{algorithm}
\usepackage{algorithmic}
\usepackage[utf8]{inputenc}
\usepackage{authblk}
\usepackage{amsmath}
\usepackage{amssymb}
\usepackage{setspace}  
\usepackage{float} 
\def\BibTeX{{\rm B\kern-.05em{\sc i\kern-.025em b}\kern-.08em
    T\kern-.1667em\lower.7ex\hbox{E}\kern-.125emX}}

\title{A Combined Feature Embedding Tools for Multi-Class Software Defect and Identification}

\author{
    Md. Fahim Sultan, Tasmin Karim, Md. Shazzad Hossain Shaon, Mohammad Wardat, Mst Shapna Akter\\
    Department of Computer Science and Engineering, Oakland University, Rochester, MI 48309, USA \\
     mdfahimsultan@oakland.edu, tasminkarim@oakland.edu,\\ shaon@oakland.edu, wardat@oakland.edu, akter@oakland.edu
}

\begin{document}
\maketitle
\begin{abstract}

In software, a vulnerability is a defect in a program that attackers might utilize to acquire unauthorized access, alter system functions, and acquire information. These vulnerabilities arise from programming faults, design flaws, incorrect setups, and a lack of security protective measures. To mitigate these vulnerabilities, regular software upgrades, code reviews, safe development techniques, and the use of security tools to find and fix problems have been important. Several ways have been delivered in recent studies to address difficulties related to software vulnerabilities. However, previous approaches have significant limitations, notably in feature embedding and precisely recognizing specific vulnerabilities. To overcome these drawbacks, we present CodeGraphNet, an experimental method that combines GraphCodeBERT and Graph Convolutional Network (GCN) approaches, where, CodeGraphNet reveals data in a high-dimensional vector space, with comparable or related properties grouped closer together. This method captures intricate relationships between features, providing for more exact identification and separation of vulnerabilities. Using this feature embedding approach, we employed four machine learning models, applying both independent testing and 10-fold cross-validation. The DeepTree model, which is a hybrid of a Decision Tree and a Neural Network, outperforms state-of-the-art approaches. In additional validation, we evaluated our model using feature embeddings from LSA, GloVe, FastText, CodeBERT and GraphCodeBERT, and found that the CodeGraphNet method presented improved vulnerability identification with 98\% of accuracy. Our model was tested on a real-time dataset to determine its capacity to handle real-world data and to focus on defect localization, which might influence future studies. The findings show significant gains in computerized vulnerability exploration, driving toward more secure software development tools.
\end{abstract}


\section{Introduction}
In this modern time, software systems become more advanced, promising their security is becoming more complicated than before. Software flaws have the potential to cause significant breaches of system security with devastating consequences for both organizations and users. Security issues in computer programs are constantly indicated through various security advisories and stored in major databases such as the CVE (Common Vulnerabilities and Exposures), the NVD (National Vulnerability Database), and so on. A review of annual CVE disclosures over the past five years reveals a notable and steady increase in reported vulnerabilities, indicating a persistent trend with no signs of decline \cite{rf1}. Over the recent years, numerous real-world attacks have been exploited software vulnerabilities, leading to significant security incidents. In 2016, Adobe Flash Player experienced substantial financial losses and a widespread compromise of user data \cite{rf2}. Furthermore, in 2019, data breaches impacting both individuals and corporations resulted in significant financial repercussions, with losses exceeding 120 and 50 million, respectively. These breaches were primarily attributed to exploits related to buffer overflow vulnerabilities \cite{rf3}. Notably, the revelation of confidential data from 533 millions individuals on Facebook in 2021 \cite{rf4}, the Lockfile ransomware attack during the same year, 2021 \cite{rf5}, the Russian cyberattack on U.S. federal agencies in 2020 \cite{rf6}, and the widespread impact of the WannaCry ransomware in 2017 \cite{rf7}. As a result, advanced methods have grown in popularity for detecting bugs earlier in the project's life cycle and preventing implementation. 

According to the study, several automated vulnerability prediction tools have been developed, including traditional approach, hybrid approach and deep-learning approach \cite{rf8, rf9, rf10}. These significantly enhancing the ability to identify patterns within source code that are probably cause the system vulnerable. However,there are several limitations in these approach. Primarily, these approaches rely on static analysis of code features to capture relevant syntactic and semantic characteristics that may provide valuable insights into potential vulnerabilities \cite{rf11}. On the other hand, some of them utilize control flow (CF) testing \cite{rf12}. This CF method often fall short to adapt to the rapidly evolving threat landscape and the advancing nature of modern software systems. Recent advances in transformers, particularly LLMs (Large Language Models), have exhibited phenomenal capability to comprehend and develop code that correlates with human syntax \cite{rf13}. However, LLM-based approaches typically struggle to capturing complex code patterns (deep nested loops, recursive calls etc), that causes a significant number of incorrect results specifically for vulnerable statements.

In this study, we address the challenge of multi-class vulnerability detection from C/C++ source code by focusing on specific CWE categories. Our approach not only identifies the presence of vulnerabilities but also locates the exact lines of code where they arise, providing developers with specific insights for remediation. The importance of this research lies in its ability to improve the detection of vulnerabilities like buffer overflows (CWE-119), memory leaks (CWE-476), and other severe flaws that threaten program security. By introducing a robust feature extraction technique, we aim to bridge the gap between conventional vulnerability scanners and the sophisticated nature of real-world software flaws. To summarize, this study highlights several major activities:

    \textbf{Feature Embedding:} We created a Transformer Graph-based feature embedding method that uses a line-sequence graph to represent the "where-the-value-comes-from" relationships between values. In this graph, each line of code is represented as a node, with edges linking successive lines to preserve them in sequential order. This structure is subsequently turned into a relational matrix, which effectively captures code dependencies and relationships.
    \textbf{Analytical Approach: } To validate performance, we compared several feature embedding methodologies, including CodeBERT, GraphCodeBERT, and our proposed CodeGraphNet, utilizing a variety of machine learning models and performance measures. The findings show that the tree-based neural network model DeepTree performed proficiently demonstrate the efficacy of our technique.
    
    \textbf{Vulnerable Lines Highlighting: }We demonstrated a vulnerable code line indicator utilizing CodeGraphNet and a prediction algorithm to reliably detect and highlight specific lines of code that could have vulnerabilities, where we presented the overall procedure as low to high, where low indicates a minor vulnerability and high indicates the severity of the vulnerability.

The remainder of this paper is organized as follows: \S2 reviews related works, and \S3 provides a motivating example. In \S4, we outline our applied approach, while \S5 details the study design. \S6 discusses the evaluation of our approach, highlighting its performance in vulnerability detection. In \S7, this works result and discussions is being summarized. Lastly, \S8 offers conclusions, limitations, and future research directions.

\section{Related Work}
Previous studies has focused extensively on improving vulnerability detection in software systems, particularly through machine learning and deep learning approaches. Where as, several studies have explored the application of feature extraction techniques with classification models for identifying vulnerabilities in code. Feature embeddings for numerical representation of source code have an important role in identifying hidden relationships inside the code, thus providing greater insight into its structure. In terms, Alenezi et al. \cite{rf14} applied word embeddings and the bag-of-words approach to automatically extract features from source code, which were fed into a neural network to refine and enhance these features. Similarly, Hovsepyan et al. introduced a word by word feature measuring techniques\cite{rf15}. However, these methods may overlook subtle, context sensitive issues within code, leading to potential misinterpretation of vulnerability pattern. Turhan et al. \cite{rf16} introduced a feature embedding technique that applies multivariate approaches, clustering similar data groups to capture interrelationships within source code. Similarly, VulSlicer \cite{rf17} provided feature embedding approaches that use program slicing from data flow graphs, which produced encouraging results. Besides, Wu et al. \cite{rf18} developed another feature extraction process based on program slicing concepts, designed to eliminate as much irrelevant information as possible, focusing only on vulnerability related aspects. However, these method filters out substantial portions of code, potentially overlooking elements that could indicate vulnerabilities. 

To fill the gaps, Devign \cite{rf19} employed graph-based neural networks to detect vulnerabilities by modeling the code as a graph structure. Similar to this,  \cite{rf20, rf21, rf22} used natural language processing methods to extract features from code samples. Although they gained good predictive performances but lacked precision in unseen set of dataset due to proper relevant feature representation. Deep learning approaches have gained more attention in addressing various software  vulnerabilities. For instance, Lam et al. \cite{rf23} proposed a CNN-based model by incorporating source code metadata alongside textual inputs, achieving improved performance. In the study by DeepFL \cite{rf24}, a recurrent neural network was combined with a multi-layer perceptron to effectively identify faults within software code. Conversely, in \cite{rf25} they employed vector representations derived from a code coverage matrix and data dependencies. Additionally, they utilized a convolutional neural network (CNN) model to predict buggy statements and methods based on the learned patterns within the data. However, a limitation of these methods is their reliance on non-contextual embeddings, which may restrict their ability to fully capture code semantics.

Regarding the transformer architecture, the integration of models like GraphCodeBERT \cite{rf26} and CodeBERT \cite{rf27} has enabled more sophisticated feature extraction, capturing both relevant semantic and structural information from source code. In LineVul \cite{rf28}, they utilize the pre-trained transformer model to capture the code-specific nuances. Transformers often struggle with capturing localized dependencies and subtle patterns in deeply nested code. Despite their self-attention mechanisms, they may overlook complex code logic (data structure, pointers call etc) and intricate variable interactions, resulting in low level pattern recognition rather than a deep understanding of code semantics. This limitation can obscure vulnerabilities and hinder precise detection.

To addresses the limitations of the previous techniques, we presented a secondary feature embedding approach termed as CodeGraphNet incorporating GraphCodeBERT with GCN (graph convolutional networks) layer. This technique adopts use of GraphCodeBERT's sense of context while also capturing complex code patterns (deep nested loops, recursive calls etc) interactions utilizing a graph-based architecture. The embedding vectors deliver beneficial insights by considering into account both local interactions and wider relationships inside the code, thereby improving the model's ability to discover possible vulnerabilities and pinpointing vulnerable line efficiently than traditional static and linear methodologies. 

\vspace{-0.2cm}
\section{A Motivating Example}
\begin{figure}[htbp]
    \hfill
    \includegraphics[width=0.50\textwidth]{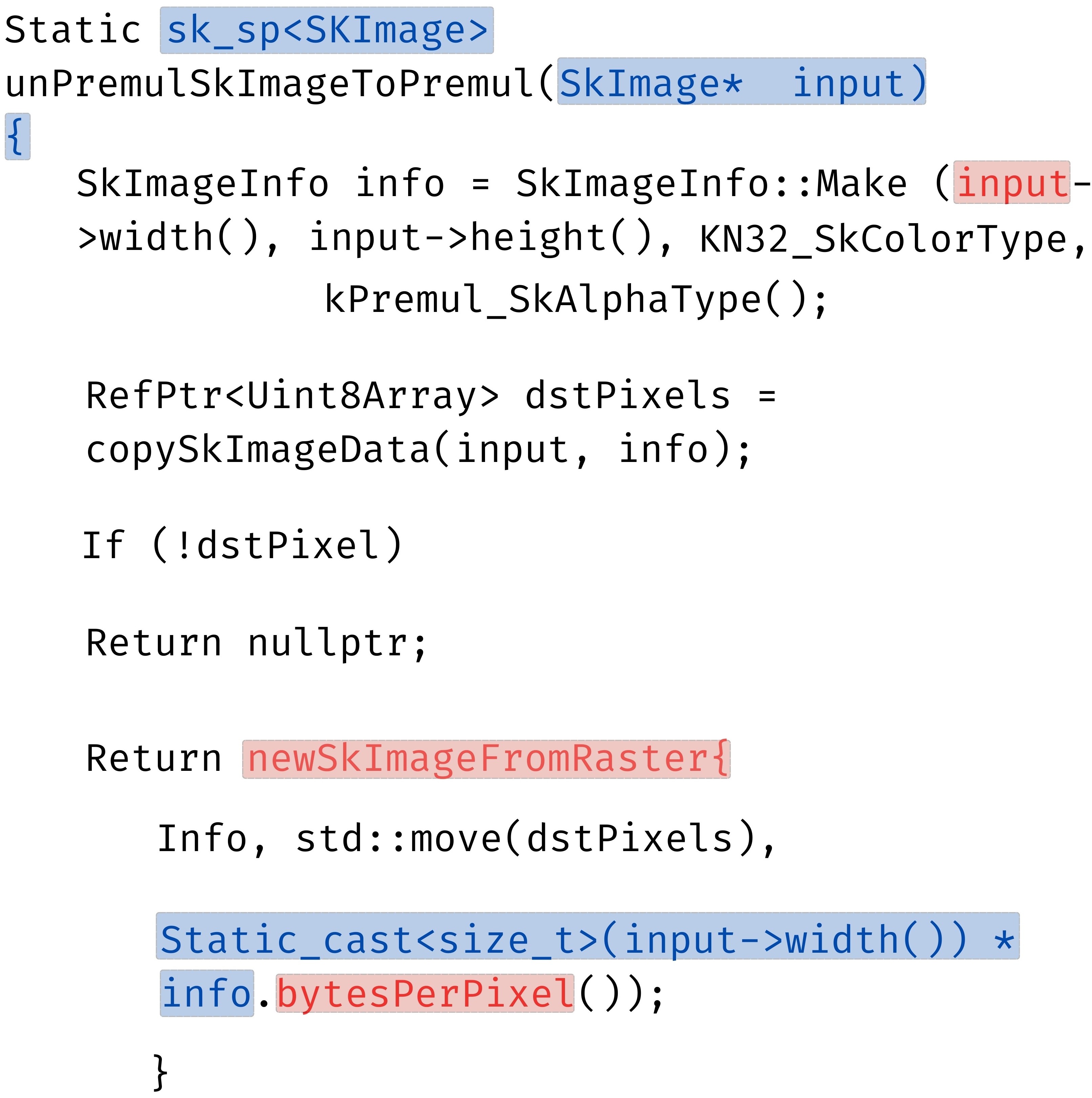} 
    \caption{Illustration of Vulnerable Code and Highlighting Potential Vulnerabilities.}
    \vspace{-0.3cm}
    \label{fig:exmp}
    \hfill
\end{figure}

In this section, we present motivating examples from our work, illustrating how traditional error localization techniques often overlook major, high-impact vulnerabilities. Here, we analyze a code snippet from AIBugHunter \cite{rf29}, as illustrated in Fig. \ref{fig:exmp}. This code example reveals two significant flaws:CWE-787 (Out-of-bounds Write) and CWE-476 (NULL Pointer Dereference). 

In Fig.\ref{fig:exmp}, AIBugHunter is represented in BLUE, while our study's findings are highlighted in RED. Although AIBugHunter identified the CWE-787 vulnerability, it overlooked the imperative CWE-476 vulnerability present in the sample. This oversight is particularly concerning, as CWE-476 can lead to severe consequences, including program crashes, undefined behavior, and potential exposure to denial of service (DoS) attacks. In terms, Our embedding model uncovers data dependencies, function calls, and contextual relationships within the code, enabling the identification of whether a specific coding error introduces vulnerabilities in other code segments. In this example, if the input is a pointer to an SkImage and it is not checked for nullptr before use, it can lead to dereferencing issues in subsequent calls. Our embeddings capture the relational attributes of input pointers in relation to other sections of code, including the SkImage pointer. This means that all potential null pointer dereferences associated with the input pointer or SkImage are identified, indicating a higher likelihood of vulnerability. Additionally, the bytesPerPixel function call is triggered by the input pointer, suggesting that the subsequent functions may also carry a significant risk of vulnerabilities. Consequently, all connected code segments lacking nullptr checks are marked in RED. In contrast, the BLUE mark at the beginning indicates a potential CWE-476 vulnerability associated with (SkImage). Additionally, at the end of the code, there exists a risk of CWE-787 if the value returned by input-width() is excessively large or negative. Multiplying such a value by info.bytesPerPixel() may lead to an integer overflow or miscalculation, resulting in an out-of-bounds write.

Overall, our work prioritizes addressing major vulnerabilities, and the error localization process is both effective and practical. By focusing on the RED indicated lines, correcting these issues can resolve not only the immediate vulnerabilities but also prevent other potential CWE vulnerabilities from arising. This comprehensive approach improves the security and stability of the code.

\section{Methodology}

\begin{figure*}[htbp]
    \centering

    \includegraphics[width=\textwidth]{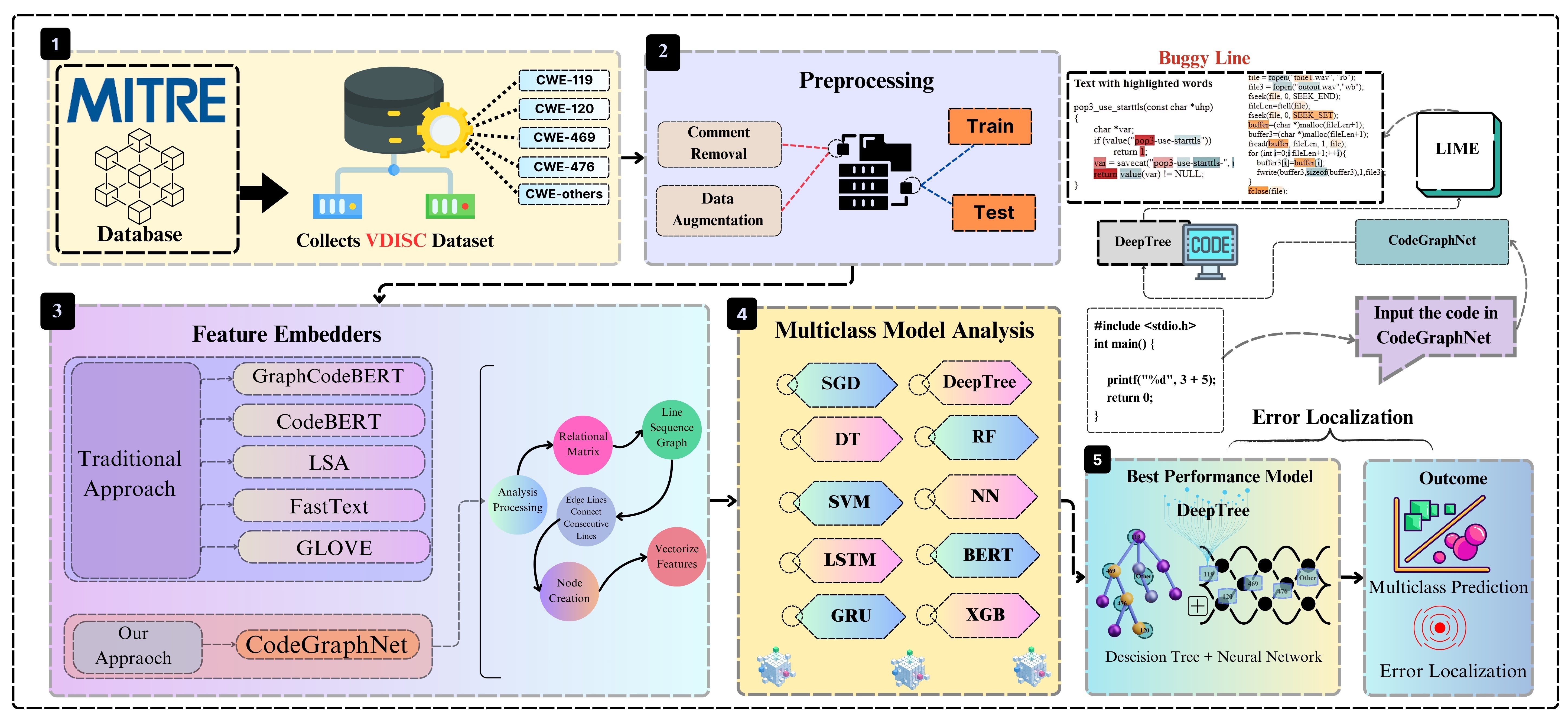} 

    \caption{Workflow of the Architecture Employed in This Study.}
    \label{fig:meth}
\end{figure*}

This section describes the entire method of our study. Fig. \ref{fig:meth} shows our comprehensive technique to recognizing various software vulnerabilities. Initially, we collected datasets from publicly available sources, particularly the MITRE Database \cite{rf30}. We obtained the VDISC dataset from these repositories, which includes CWE-119 (Improper Memory Buffer Handling), CWE-120 (Buffer Overflow), CWE-469 (Incorrect Pointer Size Calculation), CWE-476 (Null Pointer Dereference), and CWE-Others (covering non-vulnerable cases and miscellaneous instances). After acquiring the data, we converted the HD5 file to CSV format to build a concise sample which facilitated the computerized model to access proper variables and produce consistent outcomes. Despite this, by evaluating the CSV format, we discovered substantial class imbalances. To resolve this issue, we employed random balancing procedure and data augmentation approach to equalize the distribution across all distinct classes. This refined dataset was then utilized for more robust analysis and evaluation. The dataset samples used in this study are summarized in Table \ref{tab:tab1}. Furthermore, using the balanced samples, we applied both traditional feature embedding methods and our recommended embedding approaches (CodeGraphNet), along with independent testing with several algorithms and a 10-fold cross-validation strategy to determine the best performing model. When the most effective model had been selected, the study turned to error highlighting, especially determining which lines included mistakes using the LIME approach. This information will help future developers recognize and resolve issues more efficiently.

\begin{table}[htbp]
\caption{Vulnerability Classes and Data Distribution}
\begin{center}
\fontsize{10}{10}\selectfont 
\renewcommand{\arraystretch}{1.3} 

\resizebox{0.48\textwidth}{!}{
\Large
\begin{tabular}{|c|c|c|c|}
\hline
\textbf{Classes} & \textbf{Number of Data} & \textbf{Training Data} & \textbf{Testing Data} \\
\hline
CWE-119 & \Large 4502 & \Large 3605 & \Large 897 \\
\hline
CWE-120 & \Large 4496 & \Large 3557 & \Large 939 \\
\hline
CWE-469 & \Large 4500 & \Large 3608 & \Large 892 \\
\hline
CWE-476 & \Large 4503 & \Large 3632 & \Large 871 \\
\hline
CWE-other & \Large 4508 & \Large 3605 & \Large 903 \\
\hline
\textbf{Total Data} & \textbf{\Large 22,509} & \textbf{\Large 18,007} & \textbf{\Large 4,502} \\
\hline
\end{tabular}
}
\label{tab:tab1}
\end{center}
\end{table}
\vspace{-0.3cm}

\section{Study design}
In this section, we present a concise overview of the methodologies employed to strengthen vulnerability prediction in software code. Our approach begins with a feature extraction process which leads to the identification of particular code lines that indicate possible vulnerabilities.

\subsection{\textbf{Development of CodeGraphNet Feature Embedding}}
\begin{figure}[htbp]
    \hfill
    \includegraphics[width=0.49\textwidth]{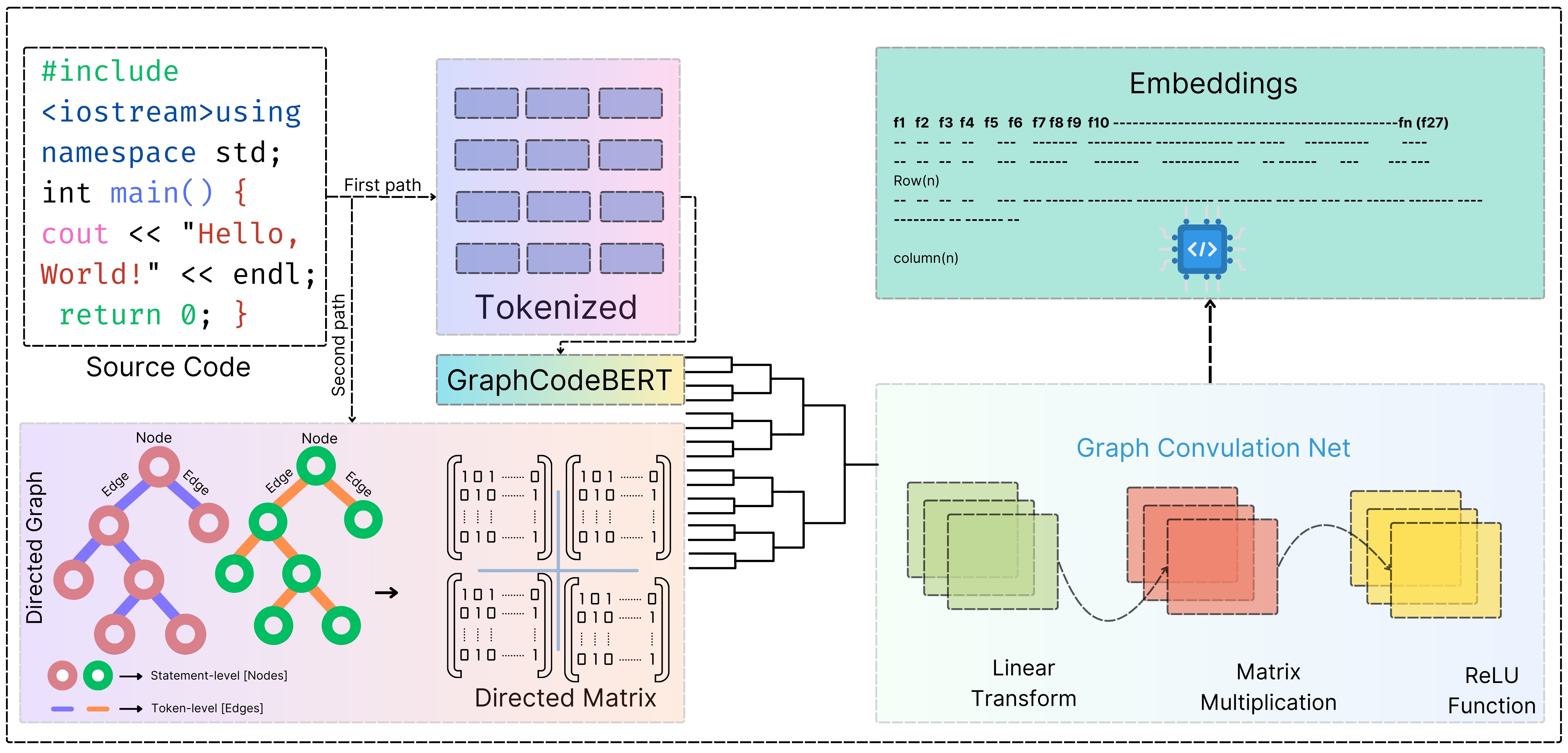} 
    \caption{An illustration of the CodeGraphNet architecture.}
    \vspace{-0.3cm}
    \label{fig:gcn}
    \hfill
\end{figure}

  We developed CodeGraphNet and the architecture of it is illustrated in Fig.~\ref{fig:gcn}. To design this approach, we have been leveraging GraphCodeBERT, a pretrained model that uses tokenization to understand code semantics and syntax. Initially, each code snippets are tokenized and processed with GraphCodeBERT, which produces embeddings for each token. The final concealed states are averaged and pooled using the formula \eqref{eq:pooling}. 
\vspace{-0.2cm}
\begin{equation}
\mathbf{h}_{\text{code}} = \frac{1}{T} \sum_{t=1}^{T} \mathbf{h}_{t} \label{eq:pooling}
\end{equation}

Here, \( T \) represents the number of tokens extracted from each code snippet, while \( \mathbf{h}_{t} \) denotes the embedding for the \( t \)-th token. Each code sample is summarized into an embedding vector, \( \mathbf{h}_{\text{code}} \), containing 768 features. To improve the accumulation of contextual relationships within code snippets, we utilize directed graphs (DiGraphs). In this structure, each node represents a line of code, and directed edges indicate the execution flow from one line to the next following the formula of \eqref{eq:graph}.

\vspace{-0.2cm}
\begin{equation}
G = (V, E) \label{eq:graph}
\end{equation}
Here, \( V \) is set of vertices primarily representing the lines of code and \( E \) is for set of edges where, \\
$V = \{0, 1, 2, \ldots, n-1\}$ \
, $E = \{(i, i+1) \mid i \in V \text{ and } i < n-1\}$

\vspace{-0.2cm}

In addition, we develop an adjacency matrix from the DiGraph, which exhibits the relationships between the nodes (lines of code) and demonstrate when one line streams into another followed by \eqref{eq:adjacency} and \eqref{eq:matrix}. 
The adjacency matrix \( A \) is defined as:
\vspace{-0.2cm}
\begin{equation}
A \in \mathbb{R}^{n \times n} \label{eq:adjacency}
\end{equation}
where \( A_{ij} \) is defined as:
\vspace{-0.2cm}
\begin{equation}
A_{ij} =
\begin{cases}
1 & \text{if } (i, j) \in E \\
0 & \text{otherwise}
\end{cases} \label{eq:matrix}
\end{equation}

This matrix provides a clear visual depiction of the code's control flow, assisting in understanding how each line interacts with and connects to the lines that follow it. These equations were used to capture the relationships and interactions among the lines. By capturing these relationships, we gain valuable insights into the structural dependencies of the code, which are crucial for effective vulnerability detection. Once the adjacency matrix is generated, the encoded representations obtained from GraphCodeBERT are fed into the Graph Convolutional Network (GCN), the equation are form in \eqref{eq:lin}.
The linear transformation of the input features \( x \) is defined as:
\vspace{-0.1cm}
\begin{equation}
x' = W \cdot \mathbf{h}_{\text{code}} + b \label{eq:lin}
\end{equation}

Using the pooled embedding from the equation of \eqref{eq:pooling},  a linear transformation is applied to adjust the feature space and dimensionality ($\mathbf{h}_{\text{code}}$) before passing it into the GCN layer. The weight matrix \( W \) contains learned parameters that scale and combine the input features, while the bias term \( b \) allows the model to adjust outputs independently of these inputs. The linear transformation outputs \( x'\). Afterwards, the GCN layer refines the initial code embeddings by propagating line level features through the adjacency matrix. 
The graph convolution step uses the adjacency matrix \( A \) to aggregate neighboring features:
\vspace{-0.2cm}
\begin{equation}
x'' = A \cdot x' \label{eq:graph_convolution}
\end{equation}

Here, \( x'' \) represents an aggregated feature that combines information from neighboring nodes, capturing the relationships between individual lines of code. The aggregated features \( x'' \) are processed through a ReLU activation function, which define non-linearity by setting any negative values in \( x'' \) to zero. This activation result, denoted as \( X''' \), improves the feature representation by applying non-linear transformations.
\vspace{-0.2cm}
\begin{equation}
(X''') = \text{ReLU}(x'') \label{eq:act}
\end{equation}

In addition, the ReLU activation function is applied element-wise to \( x'' \). Finally, this approach combines the learnt features from each line of code to produce an overall group of final features that capture both the code's structural and semantic attributes. The applied formula is:
\vspace{-0.2cm}
\begin{equation}
\mathbf{h}_{\text{final}} = \frac{1}{N} \sum_{i=1}^{N} X'''_{i} \label{eq:final}
\end{equation}

The final feature vector for each node is obtained by averaging the transformed features across all connected nodes. This extensive feature representation is essential for improving the model's capacity to properly identify vulnerabilities.

\subsection{\textbf{Comparative Feature Extractors}}
We utilized several feature extraction techniques to improve the transformation of code into vector representations. Specifically, we incorporated pre-trained LLMs such as CodeBERT and GraphCodeBERT. Additionally we used natural language (NLP) based- Latent Semantic Analysis (LSA), Fast Text Word Embedding (FastText), and Global Vector for Word Representation (GloVe) as feature extractor to capture the semantic and structural nuances of the source code. In this subsection we discuss all the feature extractor procedure.\\

\textit {B.1) LLM-based Feature Extraction}

\vspace{0.2cm}
\textbf{CodeBERT Feature Extractor}:
CodeBERT \cite{rf27} is a transformer-based model pre-trained on a vast collection of source code and natural language. The architecture of the CodeBERT is illustrated in the \textbf{supplementary file} under the \textbf{ \textit{"CodeBERT Embedding"}}.

\begin{figure}[htbp]
    \hfill
    \includegraphics[width=0.50\textwidth]{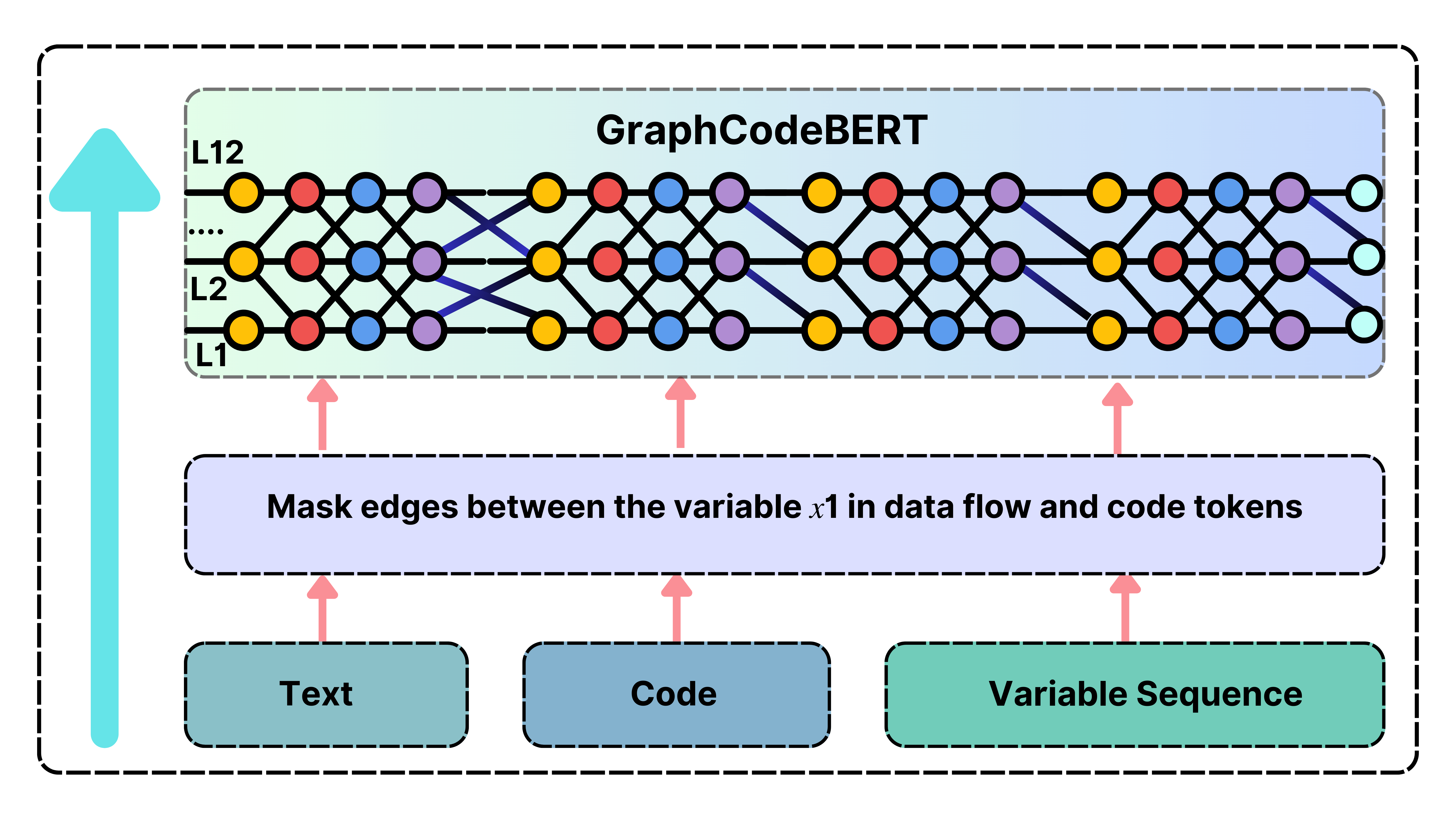} 
    \vspace{-0.7cm}
    \caption{Working strategies of GraphCodeBERT model.}
    \vspace{-0.5cm}
    \label{fig:gbrt}
    \hfill
\end{figure}

\textbf{GraphCodeBERT Feature Extractor}:
GraphCodeBERT \cite{rf26} is a powerful feature extractor specifically designed for understanding code snippets. Through the use of the underlying graph structures employed by programming languages, GraphCodeBERT is able to efficiently capture the syntactic and semantic features of the code. The working process of GraphCodeBERT depicted in Fig. \ref{fig:gbrt}.

\textit {B.2) NLP-based Feature Extraction}

\vspace{0.2cm}
\textbf{LSA Feature Extractor}: LSA effectively identifies relevant ideas underlying a symbolic significance. This method discovers latent semantic correlations between words in data while preserving the text's basic structure \cite{b40,b41, b42}. 

\vspace{0.2cm}
\textbf{FastText Feature Extractor}: FastText is an alphabetical shortening system developed by the Facebook research group in the year 2016 \cite{b38}. It comprises almost two million commonly used terms from the Practical Crawl dataset. FastText identifies itself by employing handmade n-grams as features rather than individual words \cite{b39}.

\vspace{0.2cm}
\textbf{GloVe Feature Extractor}: GloVe is a word embedding approach for converting words into dense vector representations \cite{b44}. It extracts the semantic meaning of words by evaluating global word co-occurrence data from big text corpora. GloVe assigns each word a distinct vector, with like words having near vector places \cite{b43}.

\subsection{\textbf{Classifier Model Development}}
Our technique involves constructing an efficient multi-class classifier model for vulnerability prediction. Hence, we included Stochastic Gradient Descent (SGD), Random Forest (RF), Decision Tree (DT), a proposed framework called DeepTree, Long Short-Term Memory (LSTM), Gated Recurrent Unit (GRU), Neural Networks (NN), Support Vector Machines (SVM), Extreme Gradient Boosting (XGB), and Bidirectional Encoder Representations from Transformers (BERT). This study determines and delves into the top five models that outperformed the other models. These models are extensively studied in the primary material, with a focus on classification accuracy and performance. The remaining models, including their performance and analysis, can be found in the \textbf{supplementary file} under the \textbf{ \textit{"Model Discussion"}} section for reference and comparison.\\

\textit{ C.1) Selected Best Models Description}

\vspace{0.2cm}
\textbf{Stochastic Gradient Descent (SGD)}:
SGD \cite{rf31} classifier is the most popular model. It optimizes the expected loss function through gradient descent. The SGD classifier quickly converges while handling multiple vulnerability categories. 

\textbf{Random Forest (RF)}: In this study, we utilize RF \cite{rf32} classifier for the multi-class classification to predict different types of vulnerabilities in source code. As an ensemble learning method, RF constructs multiple decision trees during training and aggregates their outputs to improve predictive accuracy.

\textbf{Decision Tree (DT)}:
In our multi-class classification task, we employed DT \cite{rf33} model due to its simplicity and interpretability. It is work by recursively splitting the data based on feature values, creating branches that lead to class labels. By utilizing this model, we aimed to classify vulnerabilities into distinct categories.

\textbf{Long Short Term Memory (LSTM) }:
LSTM \cite{rf34} networks, a specialized RNN, effectively handle multi-class classification in sequential data by capturing long-term dependencies to enhance performance. Adopting LSTMs for learning representations has become a common approach in sequence modeling.

\textbf{Large Language Model (LLM) }:
LLMs \cite{rf35} excel at multi-class classification by capturing complex patterns in text and managing diverse categories efficiently. Leveraging extensive training data and fine-tuning, it achieve precise performances.

\subsection{\textbf{Proposed Approach (DeepTree)}}
\begin{figure}[htbp]
    \hfill
    \includegraphics[width=0.50\textwidth]{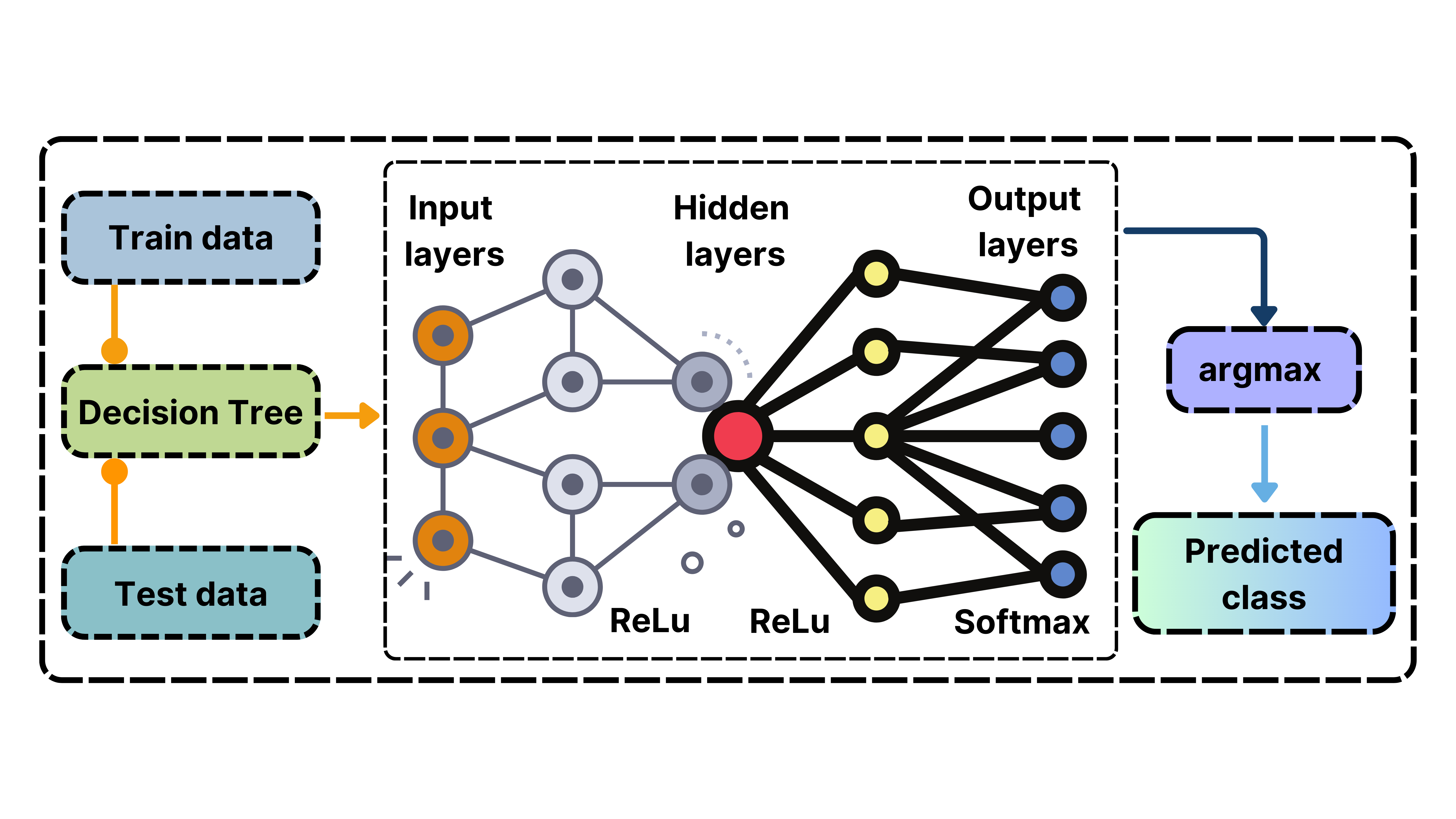} 
    \vspace{-1.3cm}
    \caption{An Insightful Overview of DeepTree's Architectural Design}
    \vspace{-0.8cm}
    \label{fig:deeptree}
    \hfill
\end{figure}

We developed a robust hybrid model that combines deep learning and traditional machine learning models to precisely detect multi-class vulnerabilities in source code termed as DeepTree. Initially, we trained a DT model on our CodeGraphNet generated dataset, which served to predict class probabilities and the predicted probabilities were subsequently utilized as transformed features for a neural network model. Overall architecture is depicted in Fig. \ref{fig:deeptree}. The model is compiled and trained using the Adam optimizer and sparse categorical cross-entropy loss function on the transformed dataset. 

\vspace{-0.7cm}
\vspace{0.7cm}

\subsection{\textbf{Error Code Highlighter}}
In this study, we incorporate LIME \cite{rf36} to develop a code line identifier that provides where is the vulnerable line. LIME provides a method to interpret individual predictions generated by models, such as DeepTree. It works by creating local approximations, perturbs the input data and monitors how DeepTree’s predictions respond to these variations. The features with the highest influence appear to be responsible for specific vulnerabilities. In contrast, LIME is capable to identifies the most influential features, enabling insights into the reasoning behind model's specific decisions by calculating weight of individuals features and the high weighted features marked as possible vulnerable influential code.

\section{Evaluation}
This section addresses key research questions aimed at evaluating the effectiveness of our proposed architecture in enhancing distinct CWE vulnerability prediction. Below are the fundamental questions regarding our study and we answer them accordingly.

\begin{itemize}
\item \textbf{RQ1. How does incorporating GCNs improve the feature embedding process for precise vulnerability detection in source code?}
    
\vspace{0.2cm}
\item \textbf{RQ2. Can the proposed model accurately identify and highlight vulnerable lines within the source code?}
\vspace{0.2cm}

\item \textbf{RQ3. How does the performance of the proposed model compare to traditional methods in CWE vulnerability prediction?}

\end{itemize}

\subsection{\textbf{Experimental Setup}}
The experiments conducted in this work were executed successfully using Google Colab Pro+ cloud based platform incorporating NVIDIA A100 GPU, Storage upto 120 GB which allows for effective handling of deep learning processes \cite{rf36}. The PyTorch deep learning framework was chosen for model implementation because of its flexibility and speed in model creation and training. Furthermore, Scikit-learn was used to create meta-classifiers. Throughout the study, this combination of tools and resources made it easier to experiment and optimize the model.

\subsection{\textbf{Performance Reports of Extractors}}

 The Table \ref{tab2} summarizes the performance of multiple machine learning models evaluated using different extraction techniques. Each row represents a distinct combination of a extractor and a machine learning model, thorough comparison of several key performance metrics, such as Area Under the Curve (AUC), accuracy, precision, recall, F1 score, Matthews Correlation Coefficient (MCC), Kappa, Mean Squared Error (MSE), Mean Absolute Error (MAE), True Positives (TP), and False Negatives (FN). \textbf{The CodeGraphNet extractor combination with the DeepTree model rivaled the others in terms of complexities. Though BERT produces excellent results, we constructed an additional feature extraction process using LLM, and Bert is another LLM model, which makes it quite challenging to build this model in the future. We chose the DeepTree model since it is simpler to construct.} This combination has the greatest AUC score of 0.97, indicating an extraordinary ability to discriminate between positive and negative classes in the dataset. This model's accuracy attained 0.98, which means that 98\% of its predictions were true. Furthermore, it achieved a high F1 score, indicating a balanced performance in terms of precision and recall. The True Positives (TP) count was also quite high, indicating that this model correctly detected the majority of positive instances, which is vital in many categorization conditions.

\begin{table*}
     \centering
    \caption{Applied Models Test Performances with Different Feature Extractors}
    \footnotesize

    \begin{tabular}{@{}llccccccccc@{}}

        \textbf{Extractors} & \textbf{Models} & \textbf{AUC} & \textbf{Acc.} & \textbf{Pre.} & \textbf{Rec.} & \textbf{F1 score} & \textbf{MCC} & \textbf{Kappa} & \textbf{MSE} & \textbf{MAE} \\ 
        \hline

        & GRU & 0.79 & 0.48 & 0.50 & 0.48 & 0.49 & 0.35 & 0.35 & 3.29 & 1.18 \\
        & NN       & 0.69 & 0.35 & 0.46 & 0.35 & 0.27 & 0.23 & 0.18 & 4.62 & 1.56 \\
         & RF       & 0.80 & 0.52 & 0.53 & 0.52 & 0.52 & 0.40 & 0.40 & 3.79 & 2.54 \\
         & SGD      & 0.74 & 0.40 & 0.52 & 0.40 & 0.34 & 0.26 & 0.25 & 4.79 & 1.79 \\

         LSA  & DT       & 0.68 & 0.39 & 0.49 & 0.39 & 0.34 & 0.27 & 0.24 & 3.55 & 1.81 \\
          & DeepTree & 0.64 & 0.41 & 0.40 & 0.41 & 0.40 & 0.26 & 0.26 & 3.45 & 1.77 \\
          & SVM      & 0.77 & 0.47 & 0.51 & 0.47 & 0.48 & 0.34 & 0.33 & 3.48 & 1.43 \\
          & LSTM     & 0.79 & 0.48 & 0.51 & 0.48 & 0.49 & 0.36 & 0.36 & 3.29 & 1.18 \\
          & XGB      & 0.79 & 0.50 & 0.50 & 0.50 & 0.50 & 0.37 & 0.37 & 3.78 & 2.14 \\
          & BERT     & 0.72 & 0.41 & 0.48 & 0.41 & 0.40 & 0.29 & 0.27 & 3.17 & 1.24 \\ \midrule

          & GRU      & 0.81 & 0.51 & 0.53 & 0.51 & 0.52 & 0.39 & 0.39 & 3.29 & 1.18 \\
          & NN       & 0.67 & 0.45 & 0.49 & 0.48 & 0.47 & 0.35 & 0.35 & \textbf{2.62} & 2.56 \\
         & RF       & 0.81 & 0.53 & 0.53 & 0.52 & 0.52 & 0.40 & 0.40 & \textbf{1.79} & \textbf{0.84} \\
          & SGD      & 0.77 & 0.47 & 0.47 & 0.47 & 0.47 & 0.34 & 0.34 & 3.17 & 1.17 \\
         Glove & DT       & 0.63 & 0.41 & 0.40 & 0.41 & 0.34 & 0.27 & 0.24 & \textbf{1.55} & \textbf{1.01} \\
          & DeepTree & 0.58 & 0.51 & 0.44 & 0.41 & 0.40 & 0.26 & 0.26 & 3.45 & 1.77 \\
          & SVM      & 0.77 & 0.48 & 0.49 & 0.48 & 0.48 & 0.35 & 0.35 & 2.18 & 1.23 \\
          & LSTM     & 0.74 & 0.58 & 0.53 & 0.48 & 0.49 & 0.36 & 0.36 & 3.29 & 1.18 \\
          & XGB      & 0.81 & 0.53 & 0.54 & 0.53 & 0.53 & 0.41 & 0.41 & \textbf{1.78} & \textbf{0.94} \\
          & BERT     & 0.65 & 0.41 & 0.40 & 0.41 & 0.40 & 0.35 & 0.34 & 2.47 & 1.54 \\ \midrule

          & GRU      & 0.77 & 0.51 & 0.50 & 0.51 & 0.50 & 0.38 & 0.38 & 3.20 & 1.13\\
          & NN       & 0.70 & 0.38 & 0.38 & 0.38 & 0.38 & 0.25 & 0.25 & 2.80 & 1.19 \\
          & RF       & 0.82 & 0.45 & 0.45 & 0.44 & 0.44 & 0.31 & 0.32 & 2.09 & 1.34 \\
          & SGD      & 0.73 & 0.45 & 0.47 & 0.43 & 0.43 & 0.31 & 0.31 & 3.24 & 1.20 \\
          FastText & DT       & 0.56 & 0.33 & 0.31 & 0.33 & 0.30 & 0.30 & 0.30 & 4.55 & 2.81 \\
          & DeepTree & 0.59 & 0.34 & 0.30 & 0.30 & 0.40 & 0.26 & 0.26 & 3.85 & 1.47 \\
          & SVM      & 0.77 & 0.48 & 0.51 & 0.51 & 0.51 & 0.39 & 0.39 & 2.18 & 0.93 \\
          & LSTM     & 0.78 & 0.48 & 0.50 & 0.48 & 0.49 & 0.34 & 0.34 & 3.45 & 1.22 \\
          & XGB      & 0.81 & 0.53 & 0.54 & 0.53 & 0.53 & 0.41 & 0.41 & 3.51 & 1.24 \\
          & BERT     & 0.72 & 0.45 & 0.45 & 0.46 & 0.40 & 0.29 & 0.27 & 2.57 & 1.34 \\ \midrule

         & GRU      & 0.84 & 0.56 & 0.57 & 0.56 & 0.57 & 0.46 & 0.46 & 3.19 & 1.18 \\
         & NN       & 0.61 & 0.45 & 0.49 & 0.35 & 0.27 & 0.23 & 0.18 & 4.62 & 1.56 \\
         & RF       & 0.74 & 0.62 & 0.55 & 0.52 & 0.52 & 0.40 & 0.40 & 3.79 & 2.54 \\
         & SGD      & 0.72 & 0.45 & 0.57 & 0.40 & 0.34 & 0.26 & 0.25 & 4.79 & 1.79 \\
         CodeBERT & DT       & 0.58 & 0.49 & 0.42 & 0.39 & 0.34 & 0.27 & 0.24 & 3.55 & 1.81 \\
         & DeepTree & 0.58 & 0.51 & 0.44 & 0.41 & 0.40 & 0.26 & 0.26 & 3.45 & 1.77 \\
         & SVM      & 0.73 & 0.57 & 0.54 & 0.47 & 0.48 & 0.34 & 0.33 & 3.48 & 1.43 \\
         & LSTM     & 0.74 & 0.58 & 0.53 & 0.48 & 0.49 & 0.36 & 0.36 & 3.29 & 1.18 \\
         & XGB      & 0.71 & 0.57 & 0.55 & 0.50 & 0.50 & 0.37 & 0.37 & 3.78 & 2.14 \\
         & BERT     & 0.69 & 0.34 & 0.30 & 0.34 & 0.30 & 0.30 & 0.29 & 2.87 & 1.26 \\ \midrule

         & GRU      & 0.85 & 0.60 & 0.61 & 0.60 & 0.61 & 0.50 & 0.50 & 2.19 & 1.08 \\
         & NN       & 0.61 & 0.52 & 0.59 & 0.52 & 0.53 & \textbf{0.41} & \textbf{0.41} & 2.96 & \textbf{1.07} \\
         & RF       & 0.83 & 0.56 & 0.55 & 0.52 & 0.52 & 0.40 & 0.40 & 3.79 & 2.54 \\
         & SGD      & 0.72 & 0.55 & 0.57 & 0.56 & 0.34 & 0.26 & 0.25 & 4.79 & 1.79 \\
         GraphCodeBERT& DT       & 0.58 & 0.49 & 0.42 & 0.39 & 0.34 & 0.27 & 0.24 & 3.55 & 1.81 \\
         & DeepTree & 0.58 & 0.51 & 0.44 & 0.41 & 0.40 & 0.26 & 0.26 & 3.45 & 1.77 \\
         & SVM      & 0.85 & 0.59 & 0.59 & 0.59 & 0.49 & 0.48 & 0.48 & 2.67 & 0.94 \\
         & LSTM     & 0.81 & 0.51 & 0.56 & 0.51 & 0.52 & 0.39 & 0.39 & 3.29 & 1.18 \\
         & XGB      & 0.85 & 0.59 & 0.60 & 0.59 & 0.60 & 0.49 & 0.49 & 1.78 & 1.14 \\
         & BERT     & 0.74 & 0.51 & 0.54 & 0.54 & 0.53 & 0.49 & 0.43 & 2.95 & \textbf{0.39} \\ \midrule

          & GRU & \textbf{0.96} & \textbf{0.95} & \textbf{0.95} & \textbf{0.94} &\textbf{ 0.94} & \textbf{0.95} & \textbf{0.95} & \textbf{0.78} &\textbf{ 0.25} \\
        & NN       & \textbf{0.94} & \textbf{0.95} & \textbf{0.89} & \textbf{0.95} &\textbf{ 0.94} & 0.23 & 0.18 & 4.62 & 1.57 \\
         & RF       & \textbf{0.86} & \textbf{0.89} & \textbf{0.89} &\textbf{ 0.90} &\textbf{ 0.89} & \textbf{0.88} & \textbf{0.90} & 3.79 & 2.54 \\
         & SGD      & \textbf{0.92} & \textbf{0.88} & \textbf{0.85} & \textbf{0.88} & \textbf{0.87} & \textbf{0.77 }& \textbf{0.83} & \textbf{0.12} & \textbf{0.19} \\
        \textbf{CodeGraphNet} & DT       & \textbf{0.83} & \textbf{0.89} & \textbf{0.87} & \textbf{0.83} & \textbf{0.89} & \textbf{0.87} & \textbf{0.79} & 2.55 & 1.81 \\
          \textbf{[Ours]} & DeepTree & \textbf{0.97} & \textbf{0.98} & \textbf{0.95 }& \textbf{0.96} & \textbf{0.96} & \textbf{0.97} & \textbf{0.95} & \textbf{1.25} & \textbf{0.77} \\
         & SVM      & \textbf{0.85} & \textbf{0.90} & \textbf{0.84 }& \textbf{0.89} & \textbf{0.90} & \textbf{0.88 }& \textbf{0.87} & \textbf{0.4}8 & \textbf{0.48} \\
         & LSTM     & \textbf{0.91 }& \textbf{0.94} & \textbf{0.93 }& \textbf{0.94} & \textbf{0.92} & \textbf{0.92} & \textbf{0.83} & \textbf{1.29} & \textbf{0.78} \\
         & XGB & 0.78 & \textbf{0.81} & \textbf{0.80} & \textbf{0.80} & \textbf{0.77} & \textbf{0.77} & \textbf{0.79} & 2.78 & 1.14 \\
         & BERT     & \textbf{0.99} &\textbf{ 0.99 }& \textbf{0.99} & \textbf{0.98} & \textbf{0.97} & \textbf{0.99} & \textbf{0.98} & \textbf{0.06} & 1.06 \\

    \end{tabular}
    \label{tab2}
\end{table*}

\begin{figure*}[htbp]
    \centering
    \includegraphics[width=\textwidth]{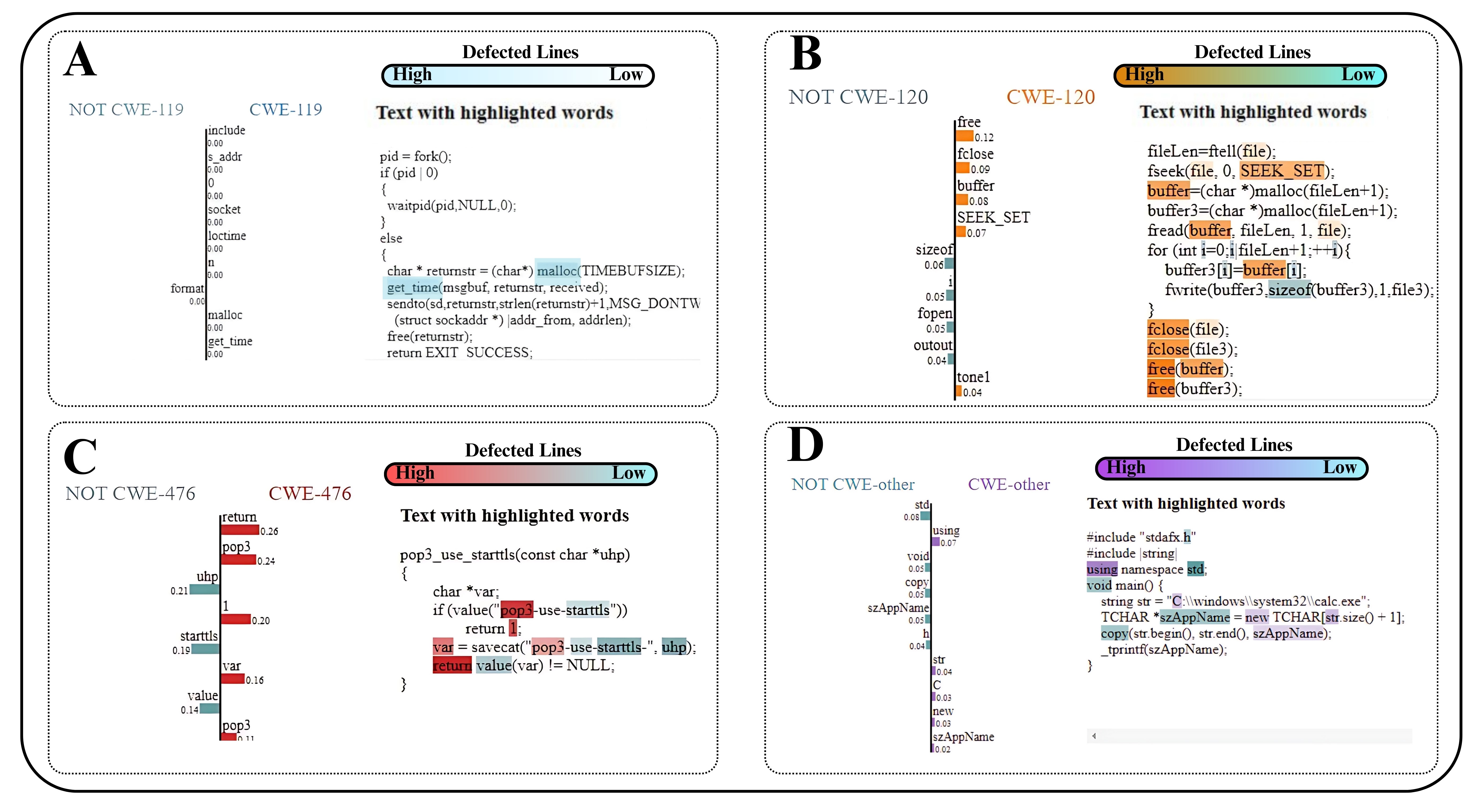} 
    \caption{Highlighting the vulnerabilities with model outcomes using LIME}
    \label{fig:ana}
\end{figure*}

In contrast, certain extractors, such as GloVe and FastText, produced competitive results, demonstrating their ability to generate effective representations for text categorization. However, it is clear that CodeGraphNet has greater feature extraction capacity, as demonstrated by its persistent high performance across many models. While the LSA extractor in particular showed poorer performance metrics across the board, which suggests that it may not be the greatest solution for this sort of categorization work. In conclusion, while numerous feature extraction approaches produced acceptable results, CodeGraphNet emerged as the most effective extractor for this classification challenge, especially when paired with the DeepTree model. This conclusion stresses the importance of feature extraction in improving model performance, and it implies that future research should focus on exploring and refining these strategies to better maximize classification outcomes. In the \textbf{supplementary file} under the \textbf{ \textit{"Performances Analysis"}}, we describe the models TP and TN, accordingly.

 We present Receiver Operating Characteristic (ROC) curves and Precision-Recall curves (PR) to illustrate the performance of the multi-class classification model. The ROC curves show the trade-off between the true positive rate (sensitivity) and the false positive rate for each class, with the AUC values ranging from 0.97 to 0.99 showed in the \textbf{supplementary file} under the \textbf{\textit{"ROC Curve and PR Curve Analysis"}} Section's. Similarly, the PR curve also shown in the same section for evaluate the model’s balance between precision and recall, with Average Precision (AP) values ranging from 0.95 to 0.98, demonstrating high precision and recall for each class.

\subsection{\textbf{Analysis of Code Line Indicator}}
To validate its robustness, we tested the approach on real-world source code containing multiple security vulnerabilities. As shown in Fig. \ref{fig:ana}, our LIME-based architecture, combined with the feature extractor and classifier model, demonstrates adequate predictions and effectively identifies the vulnerable lines in the code. In Fig. \ref{fig:ana}, deeper hues signify high impact for vulnerabilities, and lighter shades indicate minimal impacts.

\vspace{0.2cm}%
\subsubsection{\textbf{Adaptability Analysis with Stack Overflow Data}}

In this section, we evaluate the adaptability of our model using real-world code from Stack Overflow \cite{rf37}. These data reflect practical coding scenarios and common vulnerabilities.

\textbf{Improper Memory Buffer}: In Fig. \ref{fig:ana}[A], it is visible that our architecture captures the specific line of code where the malloc function is highlighted due to improper memory allocation, which could potentially lead to vulnerabilities. Furthermore, the get\_time function is also flagged, as it is involved in data handling and may contribute to the risk.

\vspace{0.2cm}%
\textbf{Buffer Overflow}: In addition, Fig. \ref{fig:ana}[B] focuses on CWE-120, where the highlighted sections indicate buffer overflow risks due to improper handling of file and buffer calls, particularly when buffer sizes are not correctly managed. The model also flags the free function.

\vspace{0.2cm}%
\textbf{Null Pointer Dereference}: Fig. \ref{fig:ana}[C] deals with CWE-476, where the highlighted segments involve variable handling (`var`, `value`, `pop3`) and return statements. These areas are potential sources of null pointer dereferences, as insufficient checks on pointer values could result in attempts to modify memory through null pointers.

\vspace{0.2cm}%
\textbf{Non-vulnerable cases}: Fig. \ref{fig:ana}[D] represents CWE-Others, such as `using namespace std`, `szAppName`, and the `str` variable. It indicates potential issues with the use of certain standard libraries. In particular, in line 6, data is copied from `str` to `szAppName`, where the string is not explicitly null-terminated. All the highlighted segments of code have been reviewed by real-world experts, addressing the concerns related to \textbf{RQ2}.

\subsection{\textbf{Validation Analysis on Unseen Dataset}}
Table \ref{tab4} presents the performance of the proposed methodology integrated with CodeGraphNet for multiclass vulnerability prediction and GrapCodeNet feature embedder, evaluated on an unseen test dataset.

\vspace{-0.16cm}%
\begin{table}[htbp]
\caption{Test Performance Results on Unseen Dataset}

\begin{center}
\fontsize{9}{12}\selectfont 
\renewcommand{\arraystretch}{1.3} 
\resizebox{0.48\textwidth}{!}{ 
\begin{tabular}{|c|c|c|c|c|}
\hline
\textbf{Classes} & \textbf{Number of Data} & \textbf{Approach} & \textbf{Acc.} & \textbf{TP / FN} \\
\hline
CWE-119 & 350 & Our Approach & 0.77 & 270 / 80 \\
\hline
CWE-120 & 350 & Our Approach & 0.87 & 305 / 45 \\
\hline
CWE-469 & 250 & Our Approach & 0.77 & 193 / 57 \\
\hline
CWE-476 & 150 & Our Approach & 0.76 & 114 / 36 \\
\hline
CWE-other & 150 & Our Approach & 0.79 & 119 / 31 \\
\hline
\end{tabular}
}
\vspace{-0.6cm}
\label{tab3}
\end{center}
\end{table}

\vspace{0.1cm}%
The model achieved varying predictive accuracies across different CWE categories, with the highest accuracy of 0.87 for CWE-120 and the lowest at 0.76 for CWE-476. Despite the model demonstrating an impressive 0.98 accuracy during training and evaluation on a separate test dataset (depicted in Table \ref{tab2}), the discrepancies in performance on unseen datasets can be attributed to several factors. Vulnerable code snippets can lead to a wide range of vital CWE vulnerabilities, and our approach identifies the most impactful ones by leveraging the specific attributes on which it has been trained. Researchers are encouraged to expand upon the architecture to achieve more robust results using a well curated dataset.

\vspace{-0.1cm}
\subsection{\textbf{Comparative Analysis}}
The present study evaluated the performance of our model with the state-of-the-art models and summarized the outcomes in Table \ref{tab4}. These models were selected due to their proven effectiveness in tasks related to code analysis and vulnerability detection, as well as their use of advanced innovative feature extraction techniques.

\begin{table}[htbp]
\caption{Comparison procedure between the present study and the existing technique}
\begin{center}
\fontsize{15}{15}\selectfont 
\renewcommand{\arraystretch}{1.5} 
\resizebox{0.48\textwidth}{!}{ 
\begin{tabular}{|c|c|c|c|c|c|}
\hline
\textbf{Models} & \textbf{Acc.} & \textbf{Prec.} & \textbf{F1} & \textbf{Code Explainer} \\
\hline
V2W-BERT \cite{rf20} & 0.94 & ---- & 0.93 & None \\
\hline
LineVul \cite{rf28} & --- & 0.97 & 0.91 & Available \\
\hline
Ref \cite{rf38}& ---- & 0.96 & 0.97 & None \\
\hline
Instruction2vec \cite{rf21} & 0.96 & 0.96 & ---- & None \\
\hline
Devign \cite{rf19} & 0.79 & ---- & 0.84 & None \\
\hline
AIBugHunter \cite{rf29} & 0.74& ---- & ---- & Available \\
\hline
\textbf{ CodeGraphNet} [This Study] & \textbf{0.98} & \textbf{0.98} & \textbf{0.98} & \textbf{Available} \\
\hline
\end{tabular}
}
\vspace{-0.6cm}
\label{tab4}
\end{center}
\end{table}

In this model comparison, CodeGraphNet appears as the most robust and reliable approach. Starting with V2W-BERT, it presents an innovative framework that enhances a BERT model for hierarchical classification of CVEs and CWEs. This model gets a high accuracy of 0.94, indicating strong overall performance; yet, it lacks F1 Score, limiting insights into its balanced performance. Similarly, LineVul employs a pre-trained BERT architecture to obtain enriched vector representations through self-attention mechanisms. Although, their code highlighting approach overlooked the imperative vulnerabilities. In, ref \cite{rf38} introduces an innovative approach to multi-class vulnerability detection that utilizes static value flow graphs alongside rich instruction embeddings within a graph neural network framework. This has an outstanding accuracy of 0.96, demonstrating success in reducing false positives. However, without accuracy score, it is difficult to assess its consistency in properly anticipating positive situations. On the other hand, Instruction2vec introduced an approach to vectorizing assembly code incorporates syntactic awareness into the learning process. Although, it exhibits harmonious performance with both accuracy and precision at 0.96. However, including with other models, the absence of F1 Score data prohibits a thorough evaluation of its precision-recall balance. In contrast, Devign, with an accuracy of 0.79 and an F1 Score of 0.84, appears less accurate but does show reasonable balance in recall and precision. Meanwhile, AIBugHunter's accuracy is noticeably insufficient at 0.74. Compared to all other models, DeepTree from this study outperformed them all, scoring 0.98 on all presented criteria, which make it excellent for accurate, dependable categorization. Furthermore, DeepTree's code is open for replication and further scrutiny, which increases its attractiveness for research and practical applications. Overall, DeepTree's balanced strengths and flexibility make it the most dependable and adaptable model in the entire range \textbf{RQ3}.

\section{Results Discussions}
\begin{figure}[htbp]
    \centering
    \includegraphics[width=0.50\textwidth]{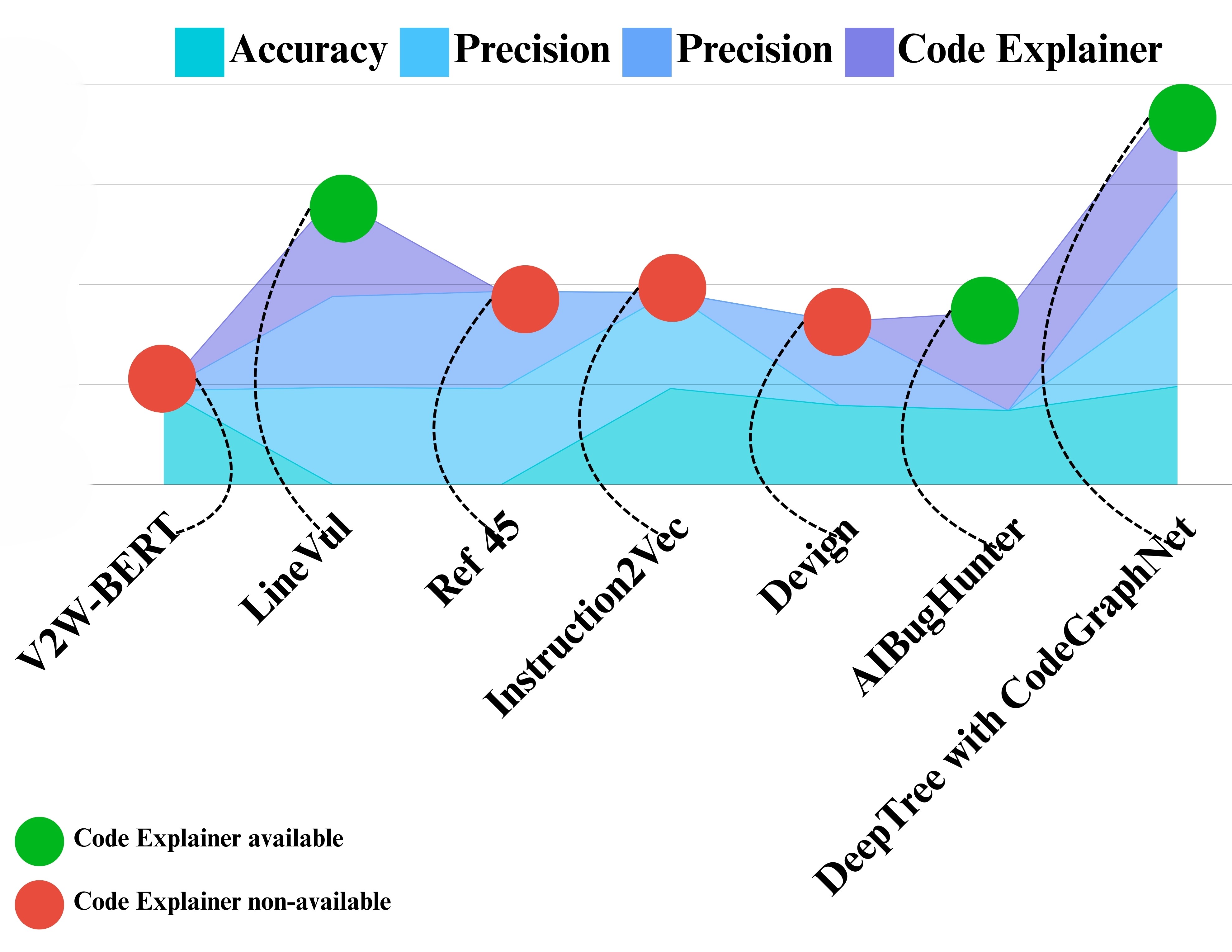} 
    \caption{Overall comparison of the model's performance with the other existing predictors}
    \label{fig:dis}
\end{figure}

In this study, we demonstrated the effectiveness of an advanced feature extraction process for vulnerability prediction, with a particular focus on highlighting vulnerable code lines. Our results clearly show that incorporating advanced feature embedding techniques (more details in \S5) significantly improves the accuracy and reliability of multi-class classification models for identifying software vulnerabilities. As seen in Table \ref{tab2}, the performance of various models improves notably when we employed our feature embedders, whereas the accuracy, precision, and overall effectiveness of the models increased. Additionally, Table \ref{tab3} presents the overall adaptability and performance improvements when both feature embedders are used in conjunction with our classifiers. However, when we integrating our multi-class classifier model with different embedding techniques, the prediction accuracy and reliability show considerable improvement, outperforming the results seen in Table \ref{tab2}. This highlights the importance of advanced classifier model in achieving better vulnerability prediction outcomes. Similarly, for the evaluation precision of our code line highlighter, we utilized real-world data from Stack Overflow, which resulted in more reliable performance across our overall architecture (see \S4). This real world validation further strengthened the effectiveness of our approach in accurately identifying vulnerable code lines. However for the better understanding we presented a visualization in Fig. \ref{fig:dis} for the comparison of the existing models and our models approaches. 

\vspace{-0.4cm}
\section{Limitations}
This study has several limitations that should be acknowledged. First, due to resource constraints, we worked with a limited dataset. Additionally, the model focuses on five distinct vulnerability classes, limiting its generalizability in broader software vulnerability detection. However, our feature extraction approach precisely identifies vulnerable code lines, it focuses primarily on detection rather than providing solutions for the vulnerabilities.

\section{THREATS TO VALIDITY}
Our study acknowledges several potential threats to both internal and external validity that could influence the reliability and applicability of our findings.

\textbf{External Validity:} Although we employed diverse code snippets from sources such as Stack Overflow, and observed reliable performances by the model. However, potential issues may arise due to overlapping between different CWEs, as vulnerabilities with similar code syntax can be challenging to distinguish. This overlap occurs when one CWE may be responsible for generating another CWE vulnerability or when they share common code patterns. As demonstrated in Fig. \ref{fig:exmp}, the same code pattern can lead to distinct CWEs, complicating the vulnerability detection process.

\textbf{Internal Validity:} In our study, several key factors emerge. First, the performance of our model is closely tied to the feature extraction techniques used, which could introduce biases due to the specific characteristics of the analyzed code. To ensure the effectiveness of the proposed feature embedders, we incorporated various feature extraction techniques to validate whether the embedders effectively capture relevant features for each sample as shown in Table \ref{tab2}.
Moreover, inconsistencies or mislabeled vulnerabilities dataset could distort the model’s learning process. In these instances, we applied preprocessing techniques to ensure the dataset was balanced and consistent, resulting in the final dataset presented in Table \ref{tab:tab1}. 

\section*{Conclusion and Future Works}
In this research, we presented AI-based approach for classifying software vulnerabilities across five classes based on graph representation combined with convolutional networks feature embedding method. Our proposed classifier, integrated with a LIME based error highlighter, effectively identifies vulnerable lines of code. Additionally, the experimental results demonstrated that our approach achieves excellent predictive performance, both in training environments and when tested on real world examples from platforms. In addition, comparative analysis with state-of-the-art models further validated the excellency of our method, showing significant improvements in accuracy and vulnerability detection capabilities. The methodology laid out in this study paves the way for future advancements in secure software development, particularly in automating vulnerability identification and providing actionable insights for code repair. In future work, we plan to refine our architecture to develop a more robust and generalized.

\section*{Data Availability}
The dataset utilized in this study and the overall steps are stored in a publicly accessible link \cite{rf39}.

\end{document}